 \documentclass[final,5p,times,authoryear]{elsarticle}

\usepackage{rotating}
\usepackage{amssymb}

\journal{High Energy Astrophysics}

\begin{document}

\begin{frontmatter}



\title{{\em Swift} discoveries of new populations of extremely long duration high energy transient}


\author{A.J. Levan}

\address{Department of Physics, University of Warwick, Coventry, CV4 7AL}

\begin{abstract}
Recent observations with {\em Swift} have begun to uncover $\gamma$-ray transients whose total energies are comparable to those of gamma-ray bursts (GRB), but have a duration an order of
magnitude or more longer than the bulk of the GRB population. Some are suggested to 
form a new population of ultra-long GRBs, with a mean duration around $10^4$s, while a further population with
$\gamma-$ray durations $>10^5$ s may represent manifestations of relativistic outflows from stars shredded around
massive black holes in tidal disruption flares (TDFs). Here I review the observations of these new classes of events, discuss
progress towards identifying their progenitors and suggest how new observations may both hone our understanding
of the outbursts, and allow them to be used as probes, that offer both complementary and additional tools to GRBs.

\end{abstract}


\end{frontmatter}


\section{Introduction}

Many of the most energetic high-energy astrophysical sources show marked variability on a range of timescales. From short duration gamma-ray bursts (GRBs) 
to outbursts from active galaxies (AGN) it is clear than the high-energy sky varies on timescales from milliseconds to many years. 
The extremes of this regime have been probed since the beginnings of X-ray and $\gamma-$ray astronomy; GRBs were first recognised in the late 1960's \citep{grb_disc}, and AGN were amongst the first identified extragalactic X-ray sources \citep{elvis78}. 
However, probing durations of hours to days has proved to 
be challenging. Such systems are too rare and fleeting to be picked up by narrow field X-ray telescopes (which are often used for studies of longer lived events) while they are also often too faint to trigger wide field rate-based $\gamma-$ray detectors, which have been 
the dominant route for mapping the rarest, but brightest events (such as GRBs). 

The {\em Swift} satellite \citep{gehrels2004} offers an ideal route to the study of transient events in these time frames. Its wide field covers roughly 1/6th of the sky in a single view of its Burst Alert Telescope (BAT). It can be triggered not only based on the rate of incident photons, but by
reconstructing an image based on a long integration, and searching this for sources that would not have triggered a rate-based detector. Indeed, it can produce integrated triggers over timescales of several days, and so long duration transients have been identified both as GRB triggers, and via an 
ongoing transient monitor \citep{krimm13}. This is coupled with the rapid X-ray and UV/optical follow-up,
meaning that 
these events, once found, can be localised and detailed multi wavelength follow-up initiated, both from ground and space-based facilities. 

\begin{figure}[h!]
\begin{centering}
\includegraphics[width=\columnwidth]{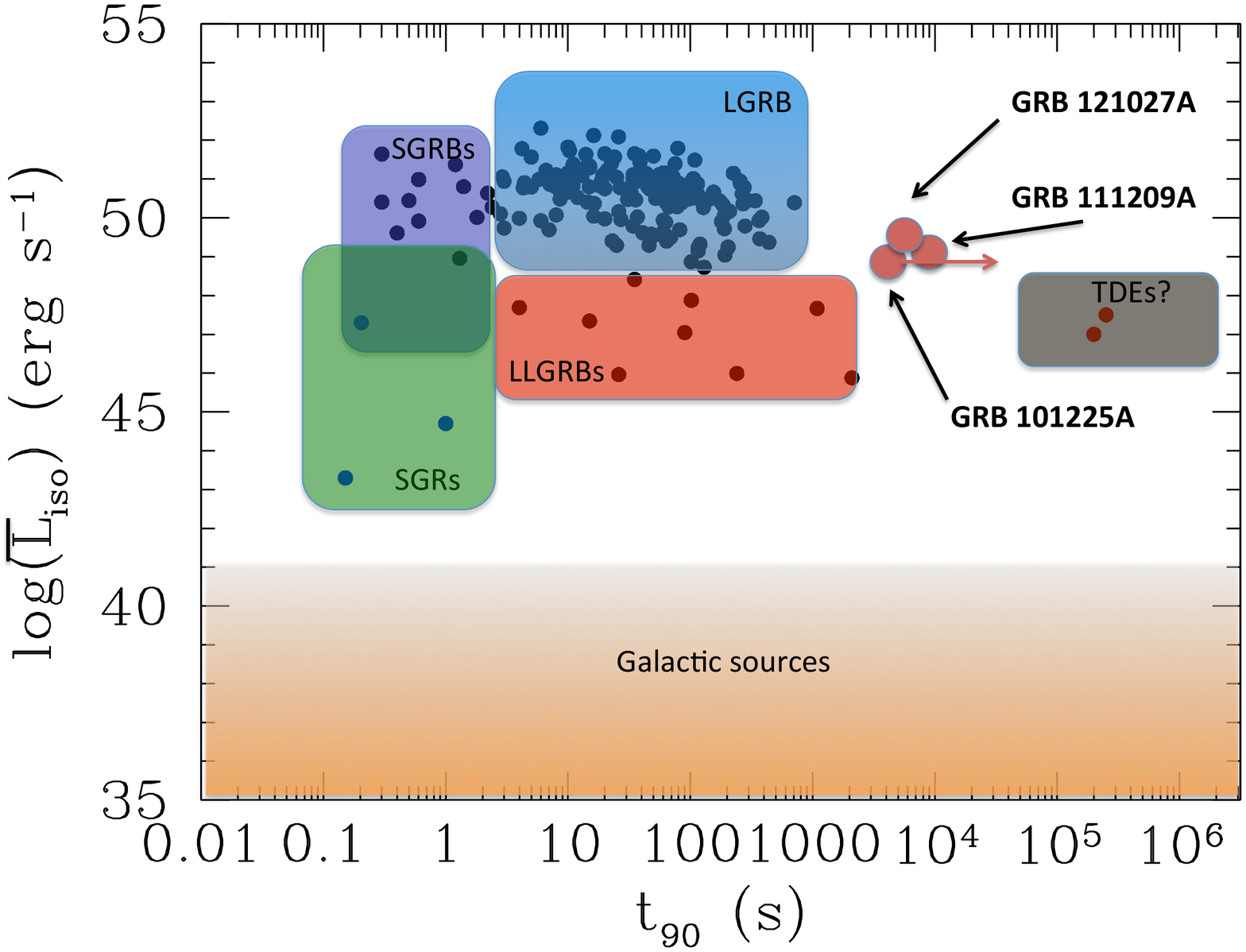}
\vskip -0.05truein
\caption{Parameter space for high energy transient phenomena, in the $T_{90}$ versus average luminosity place, from \cite{levan14}. 
Large numbers of long and short duration gamma-ray bursts can be seen, but only a handful of even longer events, lasting $10^4 - 10^6$ s. These two classes, named ULGRBs and TDFs are the subject of this review. }
\label{pspace}
\end{centering}
 \end{figure}

Remarkably, it has only been in the last few years (the second half of {\em Swift}'s lifetime, at the time of writing) that such
populations have begun to be uncovered, perhaps in part because events in the early years may have gone otherwise unrecognised, but
also because they are likely intrinsically rare. In this review I outline {\em Swift's} progress in unveiling new populations of transients that
emit $\gamma-$rays over time periods from hours to days. These duration are orders of magnitude longer than typical GRBs, but still much shorter than AGN-outbursts.  I concentrate on the most energetic events, whose total energy
releases exceed $\sim 10^{53}$ ergs. These events split broadly into two categories, with durations of $\sim 10^4$ and $>10^5$ seconds (see Figure~\ref{pspace}), they may be related, but perhaps more likely represent a diverse set of events with significantly different progenitors. For each class of
transient I will discuss the observational characteristics as a class, and proceed to consider the plausible progenitor models. There is significant overlap between their proposed progenitor systems, which break-down broadly into massive star collapses (of stars with larger radius than those creating long-GRBs), and tidal disruption flares. This in turn suggests that while the progenitors of each class may be distinct, it is possible that both classes of progenitor are present in the observed populations.

\begin{figure}[h!]
\begin{centering}
\includegraphics[width=6cm]{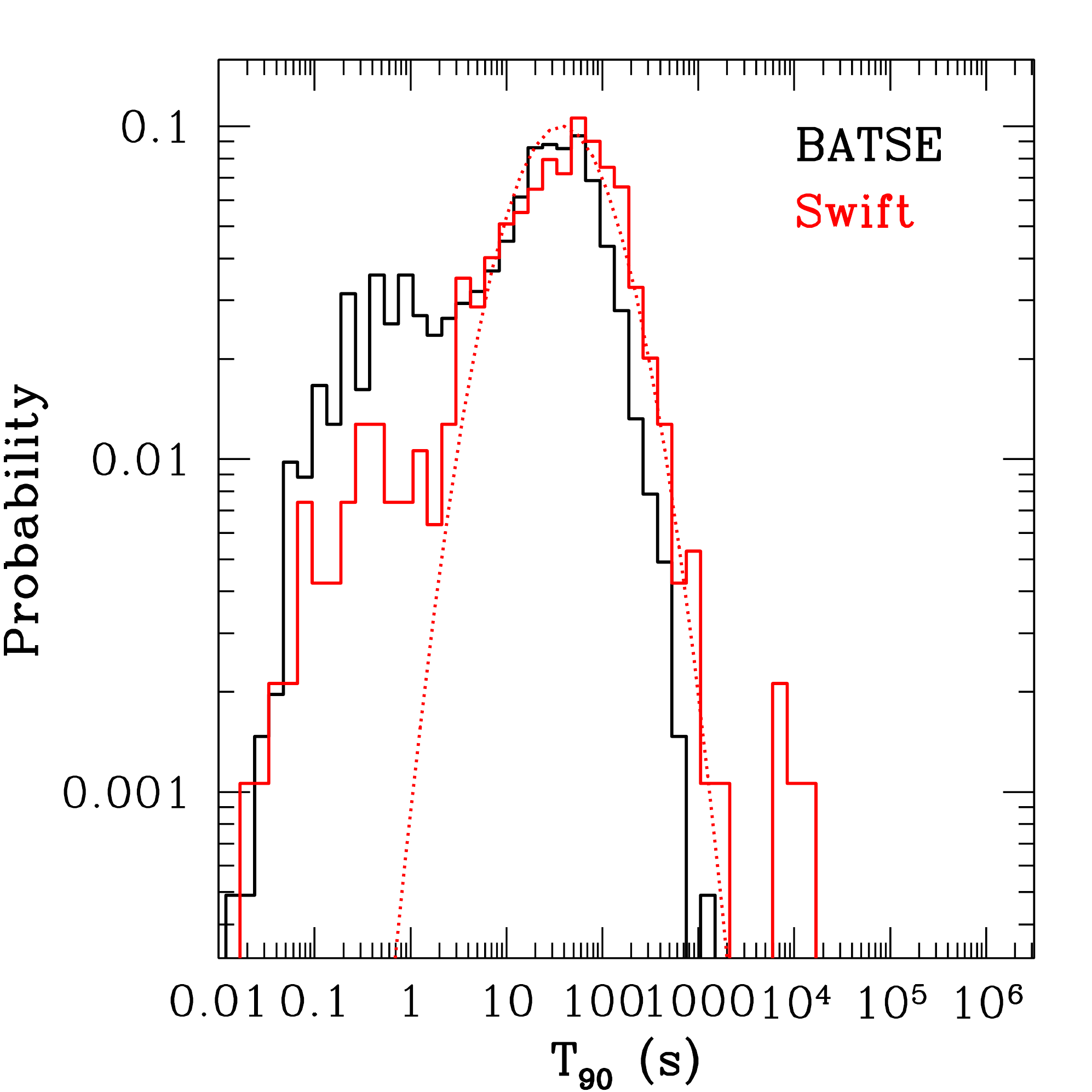}
\includegraphics[width=6cm]{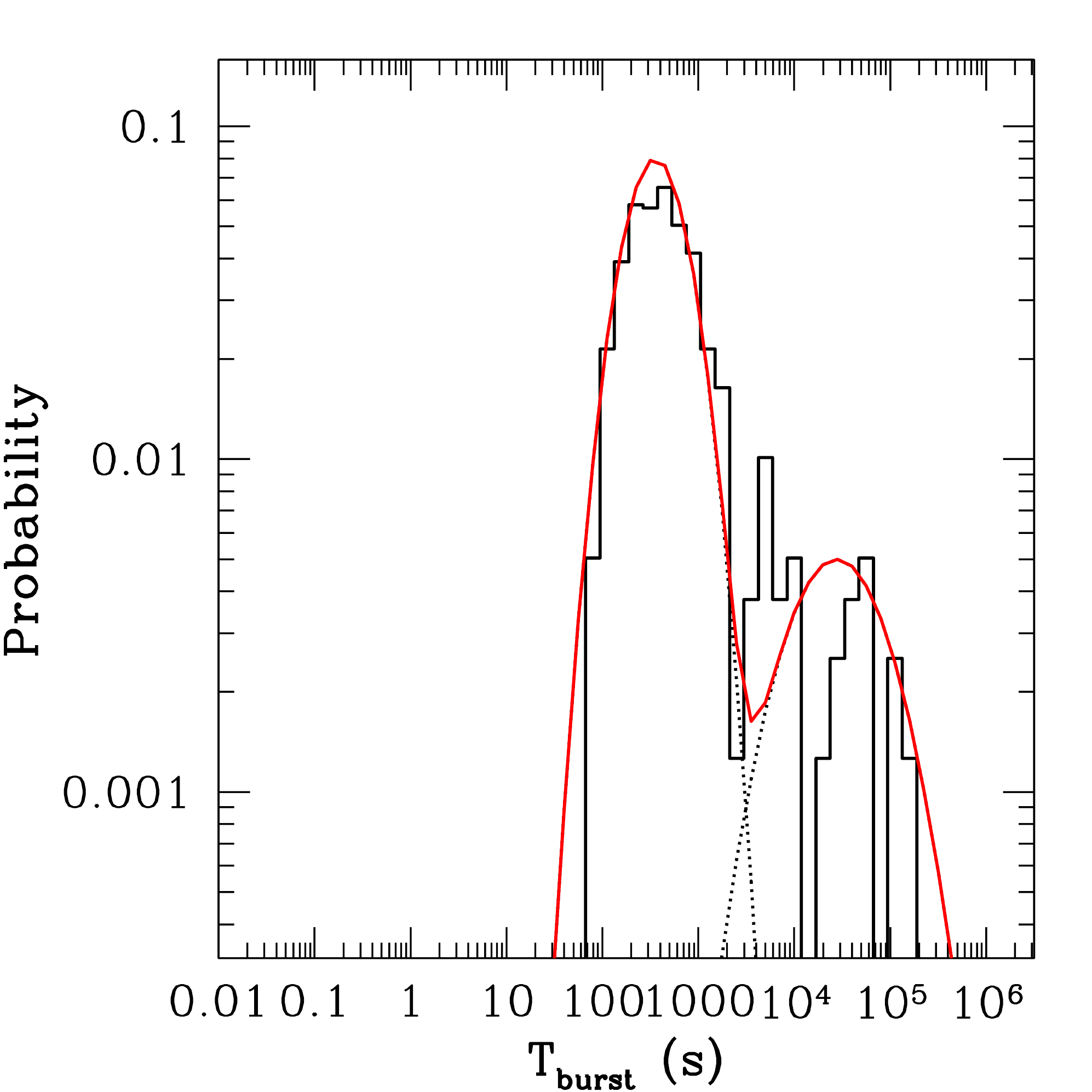}
\vskip -0.05truein
\caption{Left: The duration distribution of GRBs recorded by BATSE (black) and {\em Swift-BAT} (red). Pronounced differences
can be observed at the short end, where BATSE records many more short bursts due to its harder response. There is also 
an apparent preference for {\em Swift} to record somewhat longer durations that BATSE. The population of ultra-long
bursts are visible at the right hand side around $10^4$s. A log-normal fit to the {\em Swift} distribution from
\cite{virgili} is shown as the red dashed line. The ultra-long bursts lie well above the prediction of this model. Right: The distribution of $T_{burst}$ defined by \cite{bbzhang}, along with a two log-normal fit to the data. Typically $T_{burst} >> T_{90}$, but it is striking that in addition to a central core that is well
fit with a log-normal distribution there is a significant tail with prolonged central engine activity. }
\label{durations}
\end{centering}
 \end{figure}

\begin{figure}[h!]
\begin{centering}
\includegraphics[width=8cm]{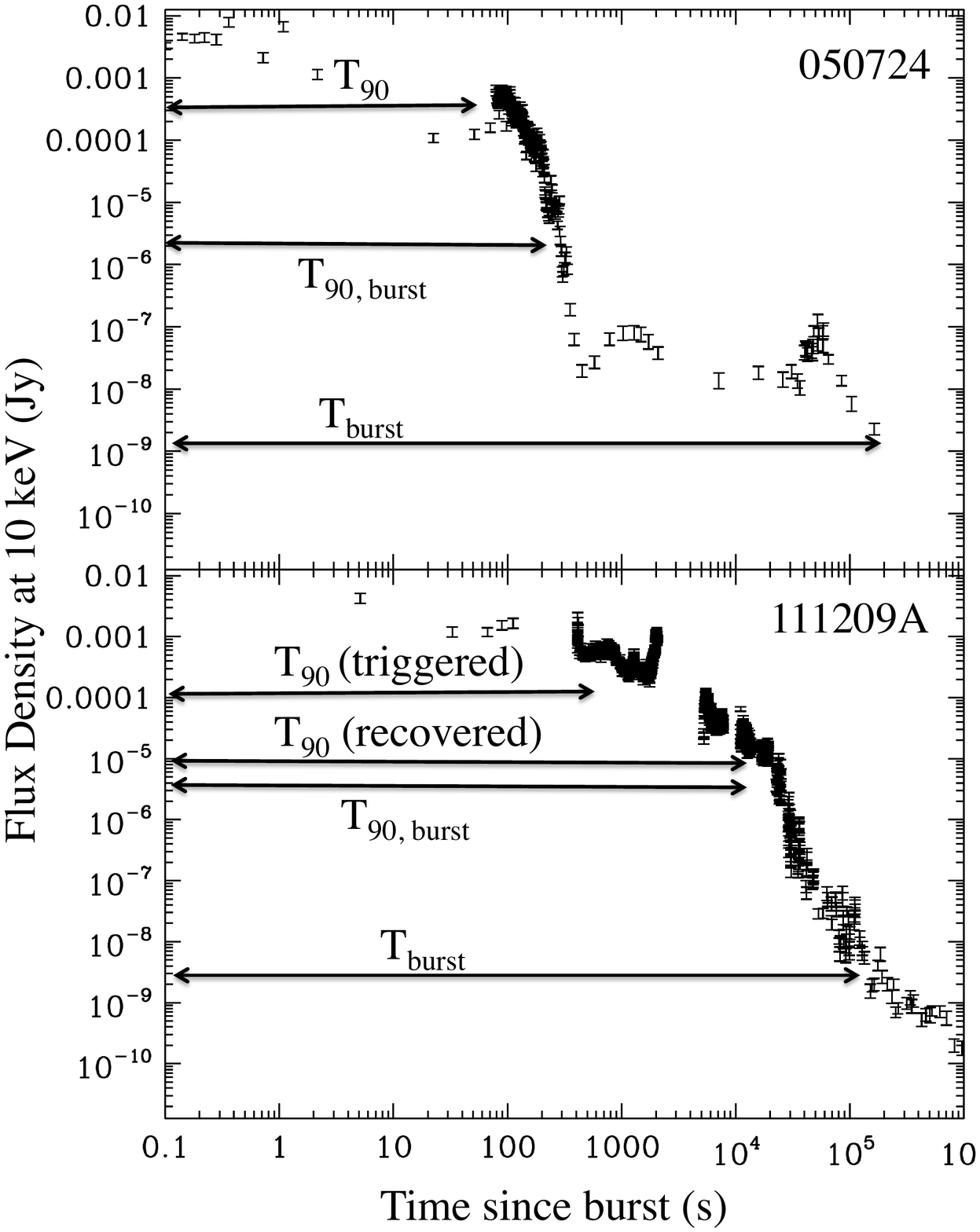}
\vskip -0.05truein
\caption{Two very different bursts highlighting the difficulty in accurately defining a duration. The upper panel shows GRB 050724, 
a burst that probably arises from the short GRB population and lies in an elliptical galaxy (SGRBs are notionally 
defined by $T_{90} < 2$\,s, although in the case of GRB 050724 in addition to a short spike there is lower-level extended
emission lasting for $\sim 100$s). While the
lower panel, shows GRB 111209A, a suggested member of the new ultra-long population of bursts. In each case different
durations have been indicated. The classical $T_{90}$ is determined by the duration of the $\gamma$-ray emission, in 
this case recorded by BAT immediately after the trigger, or in the case of GRB 111209A recovered in subsequent orbits. This
clearly misses some engine activity, and so may not be a good diagnostic. Alternatively, $T_{\mathrm{ burst}}$ defines the total activity, but
doesn't distinguish between low and high levels, and can have very different values of $T_{\mathrm{ 90,burst}}$ ($T_{90}$ determined within 
$T_{\mathrm{ burst}}$).}
\label{t90}
\end{centering}
 \end{figure}

\section{Determining a duration} 
The duration of an outburst is a fundamental parameter in its description. In addition to providing a useful tool
in the classification of a given outburst (e.g the observed distinction between short- and long duration GRBs (SGRBs and LGRBs) \citep{mazets,kouveliotou}), 
the duration of a transient also provides hints on the activity in its central engine \citep{zhang06,bromberg2}, and
has been suggested to provide direct constraints on the size of long-GRB progenitors \citep{bromberg1}. 

 Traditionally the duration of a GRB is defined as $T_{90}$, the time period over which 90\% of the
 total fluence is recorded. This is a simple and powerful definition, however it suffers from a lack 
 of comparability between instruments which may have different effective passbands (and bursts
 may last for different durations at different energies) and from the sensitivity of the instrument. A more
 sensitive instrument will follow the decay of a burst for longer. Depending on its light curve this
 could include significant additional energy, and hence increase the duration. Indeed, a comparison
 of the distribution of durations of bursts detected by {\em Swift} and {\em CGRO/BATSE} clearly shows that 
 the softer response of {\em Swift} tends to omit the typically harder short bursts. There is also a suggestion that
 the long bursts it detects have larger $T_{90}$ values, again likely because of the softer response (Figure~\ref{durations}). 
 
 In addition to the distinction between long- and short-GRBs, another feature apparent in the distribution
 of GRB durations (Figure~\ref{durations}) is the appearance of a handful of bursts with durations much longer than the majority of
 bursts observed. While very long bursts have been identified with durations of $>1000$s by BATSE, 
 BeppoSAX and Konus-WIND \citep[e.g.][]{levan05,palshin08} they have not been studied in detail, and these events 
 were generally still much shorter than some which have recently been detected by {\em Swift}
 that can be seen in 
 gamma-rays for in excess of 10,000 s -- this is the suggested population
 of ultra-long GRBs. The characteristics of this population are yet to be strongly defined, but it is clear that there is a tail of ``normal" bursts extending
 to $\sim 1000$s (see Figure~\ref{durations}), and so durations of $5000$s or more may be needed to lie clearly in the ultra-long population, we discuss this in more detail in section~\ref{ulgrb}. A further population of outbursts has been identified with durations of $>10^5$ s, these have been seen
 either as repeated GRB triggers, or via the BAT transient monitor. The durations are so extreme ($\gamma$-ray emission detected for $>10^6$s in some cases) that they clearly represent a separate population of bursts (and probably
 a separate progenitor) to the LGRBs, these very long events, candidate tidal disruption flares, are discussed in more detail in section~\ref{tde}.

\section{Ultralong gamma-ray bursts at $10^4$s}
\label{ulgrb}

\subsection{A new class? Hints from duration distribution}
Ultra-long bursts may be
 difficult to discover since long-lived, but low flux bursts are readily missed. Indeed, the {\em Swift}-BAT
 normally only records the duration as measured immediately following the trigger in so-called ``event" mode. 
 In the longest bursts the high-energy $\gamma$-rays may continue to be detected in the following orbit, although 
 it seems that with the {\em Swift}-BAT sample this is rare, with only $\sim 15$ events found in survey
 data surrounding the burst \citep{sakamoto}. Even of these only a handful exceed durations of 5000 s, and most had been 
 previously identified as candidate ultra-long events based on their prolonged, luminous X-ray emission.

 A crucial question is whether, on duration grounds alone, these bursts represent a new population of events to the traditional 
 GRBs. Different conclusions have been drawn on this. \cite{virgili} argue that 
 the overall distribution is consistent with underlying duration distribution of long-GRBs, which they find to be reasonably
 fit as a log-normal distribution with a mean of 1.47 ($T_{90} = 47$s) and
 $\sigma = 0.51$. There is clearly limited statistical power at very long durations, where the small sample sizes result in
 significant counting errors. However, they note that the 
  number of bursts observed with durations $>600$ s (11), is consistent
 with expectations of the model (9.75). However, in this case we would expect the majority of the longest bursts to
 be observed closer to this cut (i.e. with durations of 600-1000 s), in contrast, the ultra-long population suggested above
 has durations $>5000$s. This lies $>4\sigma$ from the mean inferred by \cite{virgili}, and given the
 number of bursts observed by {\em Swift} we would expect to observe none with $T_{90} > 5000$s, 
 compared with 
 $>4$ systems observed. This is shown graphically in Figure~\ref{t90}, and may indeed suggest the bursts are a new population. 
 
 Unsurprisingly, these difficulties in measuring durations have led to several authors investigating alternative 
 tools for duration determination. Of particular importance is the discovery early in the {\em Swift} mission
 that the engines of GRBs are active for much longer than the prompt emission time, with a combination
 of flares and plateaus indicating that in both long and short GRBs the engine can be active for many
 thousands of seconds beyond the end of the $\gamma-$ray emission \citep{nousek06,zhang06}. Motivated by this \cite{bbzhang} have
 attempted to define an engine lifetime ($T_{burst}$) for many of the bursts detected by {\em Swift}. This is a difficult task
 since it requires some decomposition of the afterglow light and prompt emission, both of which are likely to be contributing
within a few hundred seconds of the burst. However, the cessation of engine activity is often followed by a rapid drop-off in
brightness (formally more rapid than readily allowed by fireball blast wave models). By identifying this time it is 
possible to place some constraint on the engine activity time for 3/4 of the {\em Swift} bursts, although it is only relatively robust
in $\sim 1/2$ of the bursts. The resulting distribution is shown in the lower panel of Figure~\ref{durations}, while the outline of the duration measurement method is shown graphically in Figure~\ref{durations}. It is notable in this distribution that the
engine activity time is frequently much larger than $T_{90}$. It is also clear that while with the exception of a handful of very long bursts the
$T_{90}$ distribution is well fit by a log-normal distribution, the $T_{burst}$ distribution consists of a much larger tail, that requires
an additional component to provide a good fit \citep{bbzhang}. Indeed, a similar approach, with a similar duration distribution is reached by \cite{boer15}, who define a duration, $T_X$ to be the last rapid decay in the X-ray
lightcurve (a similar definition to $T_{burst}$). However,  \cite{bbzhang} and \cite{boer15} show less agreement over the significance and interpretation of this tail, with \cite{bbzhang} suggesting that an ultra-long population cannot be confirmed, while \cite{boer15} infer that the ultra-long population is statistically different. 
Such diagnostics are complex, the long tail 
is not necessarily representative of 
a new population since there is no strong physical motivation for the log-normal distribution of durations. Additionally, it should be noted that the rapid fall-off in the $T_{burst}$ distribution occurs at around the {\em Swift} orbital period. Given that observations of bursts cannot
normally continue much beyond the first $\sim 1000$s it is expected that this could result in a pile up of $T_{burst}$ values
on either side of this gap \citep{bbzhang}. Nonetheless, that the distributions of $T_{90}$ and $T_{burst}$ are better explained with the presence of 
an additional population at larger durations, while being well described around their peak by log-normal distributions
is suggestive of a new, additional population. 
To ascertain if these necessarily imply different progenitor systems it is necessary to turn to 
the multiwavelength afterglow properties.

\subsection{The ultra-long GRB sample} 
While there have been several examples of very long bursts over the years, the population as a whole remains poorly defined, 
and indeed is extremely difficult to define. However, the arguments above suggest that placing the duration cut towards the long
end of the bulk of the GRB distribution (i.e. in the region 500-1000 s) will result in difficulties in isolating clean samples. Furthermore, it is the
events which last much longer that are clearly distinct, and cannot be described with the same log-normal distribution that
adequately describes long-GRB durations. Therefore, conservatively one could set the ultra-long bursts as those which have $\gamma$-ray emission
detectable at the sensitivity of current instruments (INTEGRAL, Fermi-GBM, Konus-WIND, BAT, MAXI etc) of $\sim 5000$s or more\footnote{Note that by this definition, some bursts that have
previously be ascribed as ultra-long, such as GRB 020410 \citep{levan05} and GRB 090124 \citep{virgili} are not in the sample.}. 
This results in a small sample of bursts that can be considered, although it may be enlarged somewhat by the ongoing
search for further events in BAT survey data \citep{sakamoto}, or by the consideration of similar
morphological properties in X-ray light curves that might indicate bursts belonging to the same population 
\citep{levan14}. The events for which detailed mulitwavelength follow-up are
available are shown in Table~\ref{sample_ulgrb}. They are GRBs 101225A, 111209A, 121027A and 130925A. 

However, while this route of identification is robust in the sense that it identifies bursts in which the prompt emission has 
lasted for a very long time, it may also miss other members of this class. The total fluence measured for the ULGRBs above
is similar to that seen in most LGRBs, but spread over a longer time ($\sim 10^4$s versus $\sim 10^2$s), this means that these bursts are harder to detect in
$\gamma$-rays, and had they been fainter then they could easily have failed to have the longer lived $\gamma$-ray emission detected,
even if the brightest emission was sufficient to trigger the detector. Additionally, the orbit of {\em Swift} and
{\em Fermi} is such that they cannot observed the position of the burst continuously for several thousand seconds. 
Indeed, {\em Swift} rapid slewing across the sky means that it spends a few hundred seconds on average at a 
given position. These satellites may then miss some of the later time $\gamma$-ray emission. In the
case of GRB 130925A the brightest episodes as recorded by MAXI or Konus-WIND
happened at points where {\em Swift} was not observing the source, meaning that even if {\em Swift} data
is considered over several orbits, it provides a poor view of the total duration of the burst. 

Therefore it is relevant to consider if other indicators may be used as diagnostics of the ULGRB population. 
These might be \citep{levan14}
\begin{itemize}
\item Prolonged X-ray emission (lasting $\sim 10^4$ s or more) within two orders of magnitude of the peak (i.e. at the start time of the X-ray observations). 
\item A rapid decay at the end of this period, similar to those interpreted as high-lattitude emission in early GRB afterlgows. 
\item Variability within the X-ray light curve through this period that is consistent with ongoing prompt emission, but unlikely to have an afterglow origin (e.g. rapid flaring, and a strong hardness--intensity correlation). 
\end{itemize}

Many more bursts meet one or more of the criteria above, although if these are truly representative of 
the ULGRB population is less clear. Indeed, the description above is very similar to that used by
\cite{bbzhang} to define $T_{burst}$ with the additional criteria that the luminosity of the burst must 
remain within two orders of magnitude of the peak throughout this duration in order to classify 
as a ULGRB. A more quantative approach would be to utilise $T_{burst}$ as a measure of the total 
engine activity period (rather than use the total time over which $\gamma-$rays are detected), and then determine 
$T_{90,burst}$ as the time over which 90\% of the total fluence released during  $T_{burst}$ was recorded. 
In this case many GRBs would still have a short $T_{90}$, even though $T_{burst}$ was very long 
(e.g. GRB 050724, Figure 3), while because ULGRBs remain bright for longer they have a much
larger value of $T_{90,burst}$.  Further
investigations into this diagnostic may yield stronger constraints on the possible size of ULGRB population.

\subsection{Cosmological origin}
The properties of the first ULGRBs to be identified were sufficiently different from those of most GRBs that it was initially unclear if
they were of cosmological or Galactic origin, a debate very similar to that conducted for LGRBs and SGRBs in the early 1990s \citep{lamb95,pac95}. For GRB 101225A two very different models were put forward, the first that of a comet or asteroid
being tidally disrupted around a neutron star \citep{campana11}, the second the explosion of a supernovae inside a dense
envelope \citep[probably its own common envelope,][]{thoene11}. The former model indeed bares significant resemblance to 
a popular model for the creation of Galactic GRBs \citep{harwit73}, and has even recently been mooted as a possible origin for fast
radio bursts \citep[FRBs,][]{geng15}, the latter model is a rare variant on binary channels to create LGRBs \citep{izzard04,levan}, in which the explosion occurs during the common envelope via the merger of a He core with a neutron star. In this model the afterglow
evolution was reasonably fit with a type Ic supernova at $z \approx 0.3$ \citep{thoene11}

The redshift of GRB 101225A proved extremely challenging to derive, since despite a bright afterglow (see below), it was largely
featureless, showing no obvious emission or absorption lines. However, deep, later observations taken with Gemini showed
a series of faint emission lines at $z=0.85$ \citep{levan14}, resolving the Galactic versus cosmological debate, but placing
it at a redshift well beyond the model of \cite{thoene11}. Subsequently redshift measurements have been 
rather more straightforward for GRB 111209A ($z=0.667$ from both absorption lines in the afterglow and 
emission lines from the host) \citep{levan14}, GRB 121027A (z=1.71 from afterglow absorption, \citep{tanvir12}, and host emission \citep{kruhler12})  and
GRB 130925A ($z=0.35$ from the host galaxy emission lines \citep{Sudilovsky13}). Hence the ULGRB population is clearly cosmological, and to date has a rather lower mean redshift ($\bar{z} \approx 0.9$) than the LGRB population as a
whole ($\bar{z} \approx 2.24$ \citep{jakobsson06,jakobsson12}). This differing redshift is not surprising since the
lower (on average) peak fluxes of ULGRBs would naturally result in a more limited horizon.

\begin{figure}[h!]
\begin{centering}
\includegraphics[width=\columnwidth]{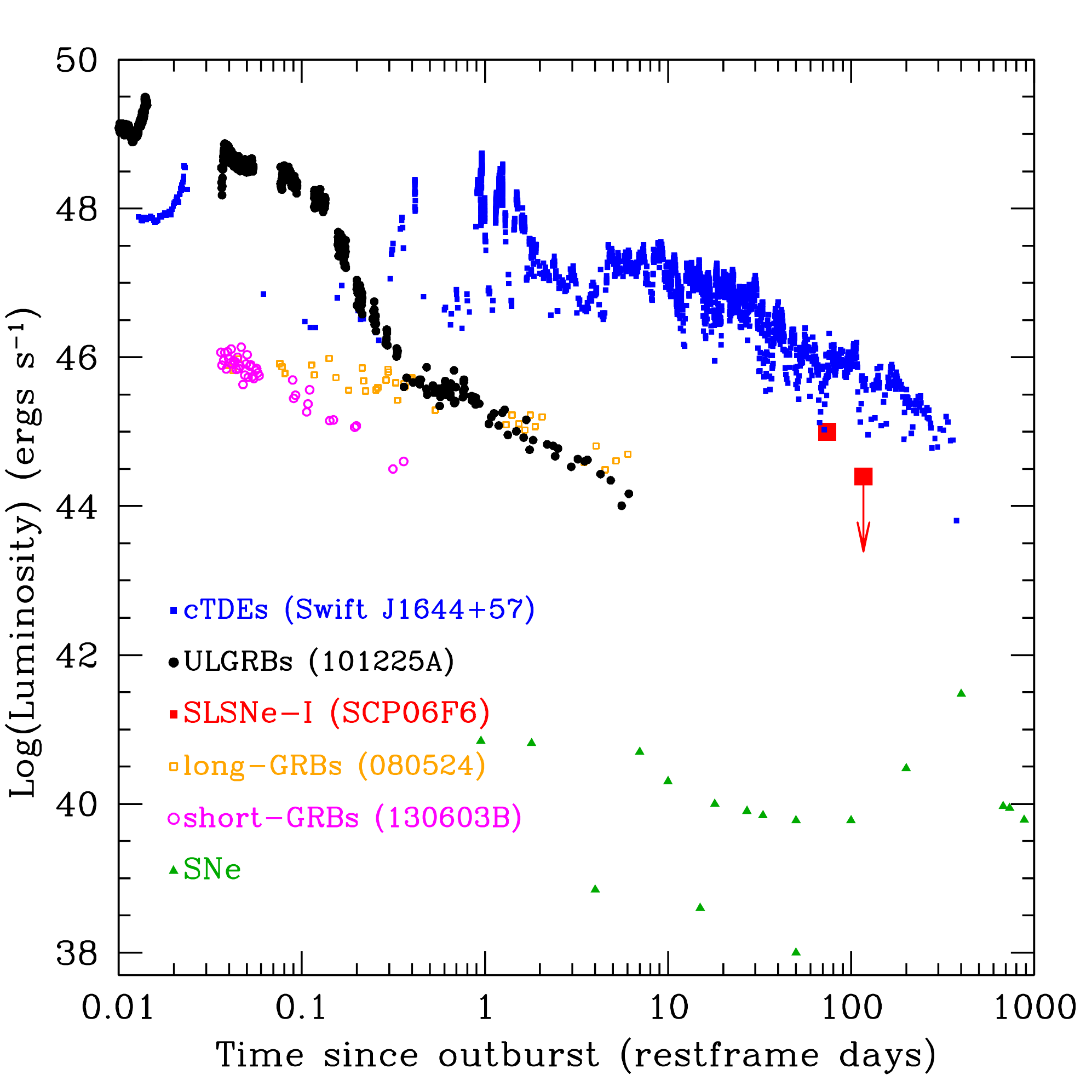}
\vskip -0.05truein
\caption{X-ray luminosity light curves of various categories of X-ray emitting transient, including supernovae, long-GRBs and short-GRBs. The new classes of very long lived and energetic transient are shown at the top of the plot. ULGRBs (black) show pronounced 
X-ray emission beyond $10^{48}$ erg s$^{-1}$ for hours following the burst, but are still an order of magnitude shorter than the
longest events (the candidate TDEs with durations of $10^5-10^6$s.  It is also interesting to note that there is one other stellar transient that
can achieve similar late time luminosities, the case of the super luminous supernovae SCP 06F6 \citep{levan13}}
\label{lc_comp}
\end{centering}
 \end{figure}

\subsection{Multiwavelength afterglow propertjes}
\subsubsection{Optical afterglows}
The afterglow properties of the ultra-long bursts form a varied set of properties, but are frequently rather different than those 
observed in the majority of long-GRBs. 
In the case of GRB 101225A, the early optical-UV light curve is exceptionally blue, and cannot be well fit by fireball models that
work well in other bursts \citep{thoene11,campana11}. In addition to this it shows marked chromatic evolution from blue to red over the course of the next few
days, again different from the largely achromatic evolution of afterglow light curves. This behaviour is
reminiscent of the evolution of the low-luminosity, also very long GRB 060218 \citep{campana06}, that is commonly
explained as due to shock breakout during a supernova. 

The afterglow of GRB 130925A is highly extinguished by dust within its host galaxy, with $E(B-V) = 2.24 \pm 0.26$ \citep{greiner14}. It can be spectrally well fit with an absorbed broken power-law, without strong evidence of spectral evolution, however the results do not appear consistent with other afterglows, and indicate the the X-rays and optical light may be originating from different locations (unlike most GRBs where both should arise from the forward shock at late times). 
Perhaps the most striking feature of the optical and IR observations of 
of GRB 130925A is that because of its extreme duration, optical and IR observations with GROND were 
possible while the $\gamma$-ray emission was ongoing. These observations appear to show the IR light curve with essentially the same structure as seen at higher energies, but with a lag of 300-400 s \citep{greiner14}. While rather different from GRB 101225A, this afterglow also therefore appears highly atypical. 

The afterglows of GRBs 111209A and 121027A on the other hand appear to have little extinction, and are well described as power-laws with indicies comparable of those found in most long-GRBs, although detailed afterglow modelling has yet to be attempted, and it is possible that the luminous X-ray emission will pose problems for combined models in these cases as well. 

\subsubsection{X-ray afterglows} 
All of the ULGRBs drive long-lived and luminous X-ray emission (see 
Figure~\ref{lc_comp} for a comparison of ULGRB X-ray light curves to other events).
At X-ray wavelengths the distinction between prompt and afterglow emission is not necessarily clear. However, 
throughout the period before the rapid X-ray decay, the ongoing flaring, and apparent hardness--intensity 
correlation \citep[e.g.][]{butler07} are consistent with the X-ray emission dominated by the prompt component. In this case the
X-ray afterglows should begin to dominate only after this decline. At this point the luminosity of the afterglows is mixed. In the case of 
GRB 101225A few X-ray's were observed after the rapid decline \citep{campana11}. For GRBs 111209A and 121027A, as with the optical 
a rather typical afterglow was observed \citep{levan14}, while in the case of GRB 130925A the late time X-rays show
strong chromatic evolution. This has been interpreted by \cite{evans14} as an afterglow which is almost absent, but with a major
contribution from dust scattered light of the earlier X-ray emission. However, in this case one would expect very little
higher energy emission, in contrast to the detection of an underlying hard component by NuSTAR \citep{bellm} that
strongly implies a significant afterglow component. However, the standard absorbed single power-law that provide a good 
description of most {\em Swift} X-ray observations fail to provide a good description of the data. These
models can work through the NuSTAR regime \citep[e.g.][]{kouveliotou13}, or can include additional hard power-law
components as sometimes seen by {\em Fermi}-LAT \citep{zhang11}. 
In the case of GRB 130925A the combined spectrum 
actually requires some additional softer component (absorption, thermal emission), whose variation compared to the power-law is probably
responsible for the apparent spectral changes seen in the {\em Swift}-XRT observations. This interpretation is also
supported by the multi wavelength data obtained by \cite{piro14}, who also find a good fit with a thermal + power law model
while also providing a fuller modelling of the afterglow. Although the intensive work on the
afterglow of GRB 130925A offers rather disparate interpretations, a common theme
of these models is that the X-ray afterglow itself is relatively weak, and consistent with being driven into a low density environment. 
Such detailed observations are not available 
for other events, but the absence of the X-ray afterglow in GRB 101225A would also support such a model. A consistently 
low density surrounding ULGRBs would provide potentially valuable constraints on their progenitors.

\subsection{Host galaxies}
The host galaxies of the ULGRB population are all star forming galaxies, with detected emission lines. 
In the case of the first two examples (GRBs 101225A and GRB 111209A) the host galaxies were apparently
extremely compact and of low luminosity (with M$_B \approx$ -16.2 and -17.5 for 101225A and 111209A respectively \citep{levan14}). 
These properties were consistent with an origin in blue compact dwarfs, with likely low metallicity, and 
such galaxies are rare, even amongst the GRB host population. In turn this was taken as some evidence 
for a distinction between the ULGRBs and the normal LGRB population. However, the hosts of GRB 121027A 
and GRB 130925A are rather brighter systems. 

A further important diagnostic is the location of the events within the host galaxies. Long-GRBs are highly
concentrated on their host light (much more so than normal core collapse SNe, \cite{fruchter,svensson}), while short-GRBs are
highly scattered at large offsets \citep{fong10,fong13}. 
Again, the first two examples of ULGRBs showed an apparent origin consistent with the nuclei of
their hosts, perhaps indicative of an origin with the black hole that may reside there. However, these
small galaxies also contain a significant fraction of their total light within the brightest pixel of their hosts (since
they are compact), and so the probability of association with this location should they follow a distribution
similar to GRBs is significant. Indeed, the case of GRB 130925A clearly shows that not all events are nuclear, 
with {\em HST} imaging clearly demonstrating an non-nuclear origin \citep{tanvir13}. However, in this case
the galaxy also shows signs of recent merger activity, and so may harbour more than one massive black hole.

\subsection{Supernova searches}
Three out of the four well studied ULGRBs lie at redshift where supernovae could be visible should they
be similar to those seen in normal LGRBs \citep{li14}. Notionally, this should be most straightforward in
the case of GRB 130925A, whose redshift $(z=0.35)$ is almost identical to the case of GRB 130427A, whose
SNe was readily identified \citep{xu13,levan14}. However, the burst afterglow was apparently highly extinguished
\citep{greiner14}, 
and only seen in the near-IR, where detailed follow-up was more difficult. To date there is no 
clear evidence for an SNe in this case. 

This leaves the cases of GRB 101225A and 111209A. Both of these events show strong UV emission, and
little evidence for significant host extinction. Despite their higher redshifts they are promising 
routes to searching for associated SNe, as deep, late-time multicolour observations were taken in each
case \citep{thoene11,levan14}. In both cases there is spectral evolution, with a change from blue $\rightarrow$ red. This could well be the hallmark of supernovae emission. GRB afterglows are typically 
described as power laws (modified by host extinction, or the unusual possibility of a spectral break in the range of
interest), in contrast SNe exhibit strong metal line blanketing short ward of $\sim 3000$\AA~ in the rest-frame. 
Hence the emergence of an associated SNe should provide a measurable reddening, and may well reverse the decay
at rest-frame optical wavelengths where the SNe peaks. 

\begin{figure}[h!]
\begin{centering}
\includegraphics[width=\columnwidth]{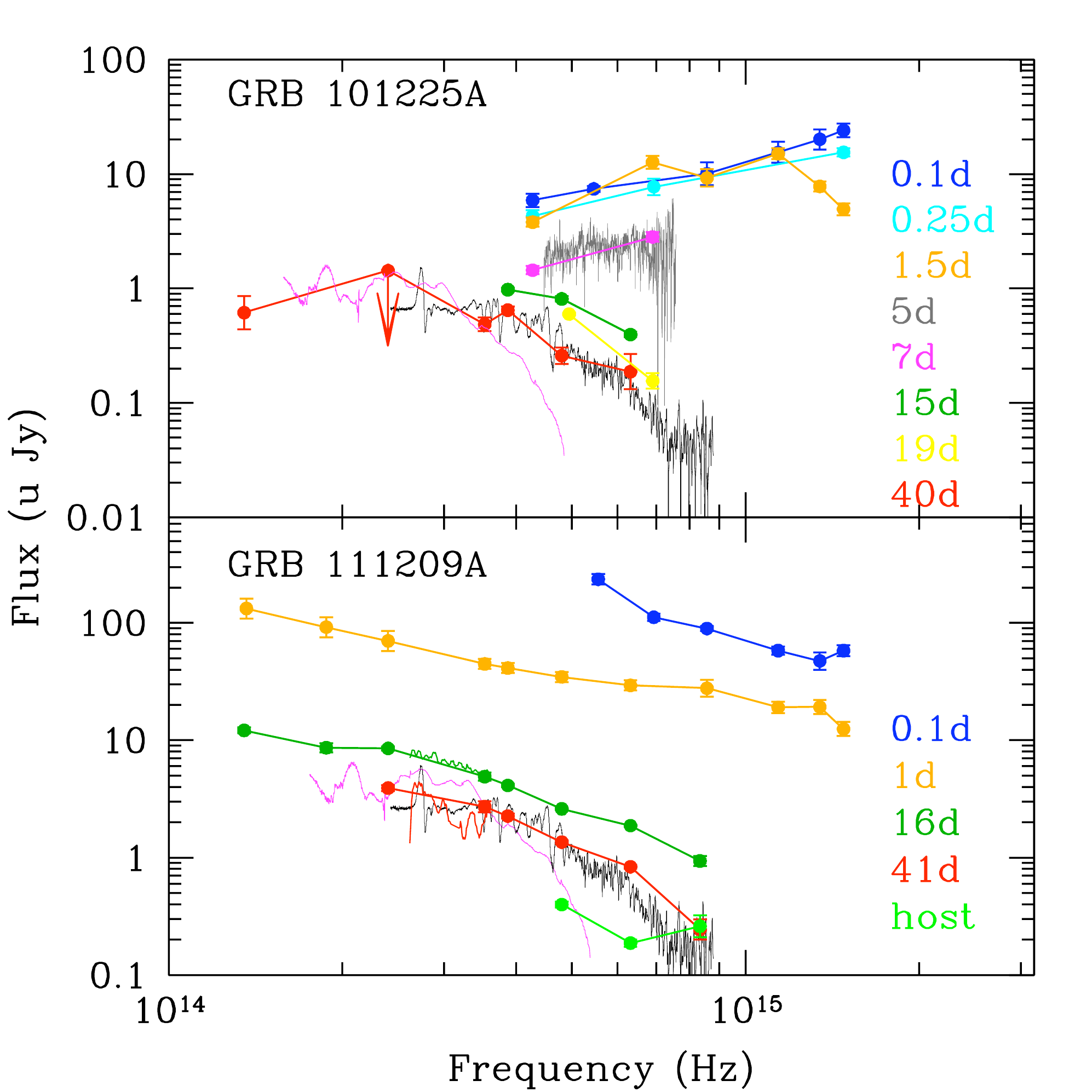}
\vskip -0.05truein
\caption{The evolution of the spectral energy distribution of GRB 101225A and GRB 111209A (from Levan et al. 2014). Both
can be seen to evolve from blue to red, with the initial SED of GRB 101225A rising into the blue (in contrast to GRB
afterglows that scale typical as $\nu^{-1}$). In broad terms this evolution is what is expected from emerging SNe. However, 
the counterparts remain too blue to be well fit with simple SN templates similar to those seen in normal GRBs. }
\label{sed_evolve}
\end{centering}
 \end{figure}

\begin{figure}[h!]
\begin{centering}
\includegraphics[width=\columnwidth]{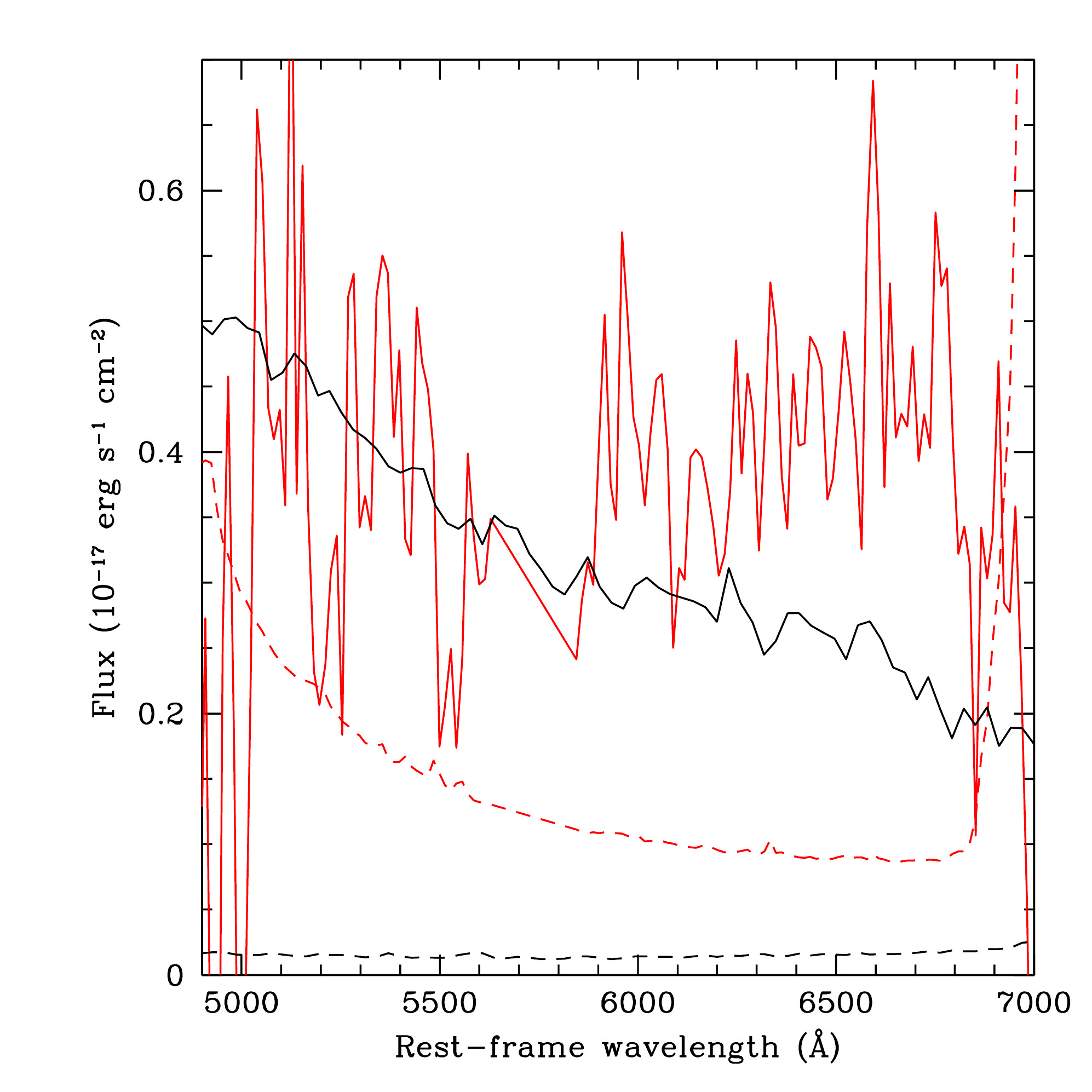}
\vskip -0.05truein
\caption{{\em HST} grism spectroscopy of the long-GRB 130427A \citep{levan14b} and
the ultra-long GRB 111209A \citep{levan14}. The flux levels have been scaled to
match for comparison of the overall spectral shape, and have not been 
corrected for either afterglow or host galaxy contribution. The case of GRB 130427A clearly
shows features consistent with a SN~1998bw SNe (in particular a broad bump at around 5000\AA). In
contrast, no such features are seen in the lower signal to noise spectrum of GRB 111209A. However, 
despite the signal to noise it is apparent that the overall spectral shape in GRB 111209A is much flatter
(in $F_{\lambda}$) than expected for an SNe. This could be due to a larger contribution from the
host galaxy, because the SNe type is different (for example the flatness in this spectral regime may
be due to enhanced emission around 6500\AA from a broad H$\alpha$ in a SN II), or because
there is no SNe in this ULGRB. }
\label{grism}
\end{centering}
 \end{figure}

However, while the first order properties seen in GRB 101225A and GRB 111209A appear to support this model, 
they differ in the details. In particular, the counterparts remain too blue to be well described by a
supernovae similar to the SN~1998bw models that provide such a good fit to most LGRBs (see Figures~\ref{sed_evolve}) and spectroscopy taken of GRB 111209A at the time of expected SNe peak fails to show any obvious SNe features (Figure~\ref{grism}).

\begin{table*}[htdp]
\begin{center}
\begin{tabular}{llllllll}
\hline
GRB &  $T_{90}$ (s) &  $T_{burst}$ (s)  & $T_{90,burst}$ (s) & $z$ & E$_{iso}$ (erg) & Host mag &  References  \\ 
\hline 
101225A  & 7000 & 106659 & 21800 & 0.85  & $5.2 \times 10^{52}$ &  -16.2 (g) & \cite{levan14,thoene11} \\
111209A  & 13000 & 63533 & 14300 & 0.677 & $7.0 \times 10^{52}$ &  -17.5 (r) &  \cite{levan14,gendre13}\\ 
121027A   & 6000 & 35399 & 10700 & 1.773 &  $8.8 \times 10^{52}$  & N/A &   \cite{levan14,starling} \\
130925A   & 5000 & 11614 & 2700* & 0.35 & $7.0 \times 10^{51}$& -18.4 (g) &  \cite{evans14,piro14}  \\
\hline
\end{tabular}
\end{center}
\caption{The {\em Swift} sample of ultra-long GRBs with detailed X-ray, optical and IR follow-up. $T_{90}$ is 
based approximately on the duration over which $\gamma$-ray emission was detectable to BAT, but is different from
that reported by standard analysis due to the continuing contribution over subsequent orbits. $T_{burst}$ is taken
from \cite{bbzhang}, and $T_{90,burst}$ is approximately estimated based on the cumulative flux density with
$T_{burst}$ (light cures following \cite{evans09,evans10}. The redshifts given are from the literature, while the isotropic energy releases ($E_{iso}$) are based
on the integrated 10 keV flux density, and do not include any 
correction to a wider band. *In the case of GRB 130925A the brightest emission was missed by BAT, and so the
integration yields a smaller $T_{90,burst}$. The host absolute magnitudes are given simply by $M = m - 5 \log(d/ 10 pc)) + 2.5 \log(1+z)$, and
are not further k-corrected. The brackets refer to the band of measurement. }
\label{sample_ulgrb}
\end{table*}%

\subsection{Progenitor models} 
There are various models posited for the origin of ULGRBs, each have distinct advantages and constraints when 
describing their observed properties. To date there is no smoking-gun observation of the progenitor of an ULGRB, and so
each of these models remain plausible. 

The observed duration of a GRB  ($T_{\gamma}$) is dependent on the time that the central engine is active ($T_e$), minus the
time that the jet takes to penetrate whatever dense medium surrounds the central engine ($T_b$), i.e.  $T_{\gamma} = T_e - T_b$ \citep{bromberg1}. Ultra-long bursts then require pro-longed activity of the central engine. For most 
GRBs it seems likely that the progenitors are very compact, leading to small breakout times.  Fine tuning issues would
be problematic in creating short periods of $\gamma$-ray emission if the time to breakout was much longer  (i.e. the scenario where $T_e \approx T_b >> T_{\gamma}$ would require fine tuning). For ULGRBs such constraints
are not present, and it is possible, but by no means required, that much longer breakout times (corresponding to 
much larger radii progenitors) can create ULGRBs. 

\subsubsection{Collapsars}
Collapsars are appealing as the progenitors of ULGRBs since they are known to produce long duration GRBs and
so can clearly create events with the necessary energy, spectra etc. It is less clear how they may create such 
long duration events, and what may distinguish the longest duration events from ``normal" LGRBs. In most
GRBs the long-lived engine is thought to be powered by fall-back from material expelled in the burst, but with
insufficient velocity to escape the newly formed black hole \citep[e.g.][]{zhang06,fryer07}, or alternatively from material that fragments
during the collapse process \citep{king05}. In the former case one would expect the infall rate to reduce
with time (naively the fallback accretion rate scales as $\dot{M} \propto t^{-5/3}$ \citep[e.g.][]{fryer06,fryer07}). In the latter
case one might expect flares, rather than continuous emission. This doesn't easily produce the 
observed light curve properties of ULGRBs. 

However, the observed $\gamma$-rays are not just the product of the central engine, but 
(at least in the case of LGRBs) the interaction of the shocks that are produced in the outflow from this engine. 
These shocks decelerate in the medium surrounding the burst, and then shock with it to create the
afterglow. It has been suggested in the case of GRB 130925A (and possibly applicable to
other ULGRBs) that a low density medium surrounding the GRB would result in a significantly longer
deceleration time \citep{evans14}, with internal shocks between ejected shells continuing to 
produce $\gamma$-rays over this period. 
The challenge for this scenario is that the low density medium would be expected to drive only 
a weak afterglow. In the case of GRB 130925A this is side-stepped since it appears to be in a dusty
environment, and the unusual afterglow that softens with time may be explained by a
dust echo of the prompt emission \citep{evans14}. It may also not represent a problem for GRB 101225A, where there
is essentially no X-ray emission beyond the early rapid decay (and so potentially no afterglow). However, in the case of GRB 111209A and
GRB 121027A, these is little evidence for significant dust extinction, and the afterglows appear much more typical \citep{levan14}. 

Finally, the absence of obvious SNe signatures in the afterglow of GRB 111209A would appear to 
conflict with this model. With two exceptions, that may belong to a short-population of bursts \citep[e.g.][]{fynbo06}, LGRBs are
known to be associated with high velocity type Ic supernovae. Indeed, recent work shows these to 
be remarkably uniform, and may even enable their use as standard candles \citep{li14}. However, 
in the case of GRB 111209A there is no such signature, despite deep {\em HST} observations that should
have readily detected it \citep{levan14} (see Figure~\ref{sed_evolve}). This implies that this burst was not a typical LGRB, and hence its progenitor
may well have not been a typical collapsar.

\subsubsection{Collapse of giants or supergiants}
The collapse of giant stars offers an alternative scenario to explain ULGRBs \citep{gendre13,levan14}. These systems are known to create type IIP SNe based on the direct detection of their progenitors \citep[e.g.][]{smartt09}, but their ability to create high-energy transients remains unclear. 
While collapsar progenitors typically have radii of only a few $R_{\odot}$, supergiants have radii of an AU or greater. Their collapse
times are therefore orders of magnitude larger than in collapsars, and given their lower densities their $\dot{M}$ rates are also
smaller. The fuelling of the nascent black hole would then naturally create a burst of extremely long duration. However, 
it is less clear if the collapse would necessarily create the necessary accretion disc geometry to form a GRB-engine, or
if the Eddington limit should start to apply during the prolonged collapse, essentially halting the infall and stalling the GRB. However, numerical simulations \citep{nakauchi13} do appear able to drive transients with the requisite durations from
giant star collapses.

GRB 130925A offered an ideal opportunity to test these models, which were largely developed as a result
of observations of the earlier bursts. Unfortunately 
optical extinction from the host makes a direct search for SNe II impossible, however, harder X-ray observations do not suffer from the same problem and have 
been exploited to this end. While \cite{evans14} interpret the evolution of the {\em Swift}-XRT data as
due to an evolving dust echo, the combined {\em Swift}, {\em XMM-Newton} and {\em NuSTAR} observations
are well fit with a power-law plus a thermal component. \cite{bellm} assume this thermal component
to originate in a hypercritically accreting disc, whose luminosity is consistent with being fed by 
fallback from a supergiant star (although the range of allow fallback rates is large enough to encompass many models
\citep{wong14}) . 
\cite{piro14} advocate an alternative interpretation in which the thermal component arises from cocoon emission
surrounding the relativistic jet, and importantly note the requirement of a low ambient density, consistent with one sculpted by 
a blue supergiant star at low metallicity. 

While it remains unclear if this model will survive further scrutiny (for example the the
host region appears to be at high metallicity \citep{schady15}, and the robustness and uniqueness of the thermal solution is unclear \citep{evans14}), 
future observations both in the X-ray and at optical wavelengths should provide direct tests of giant and supergiant models.

\subsubsection{Tidal disruption flares}
An alternative model for the origin of ULGRBs is that they arise from tidal disruption events. The appeal of this model
is two-fold. Firstly, a tidal model has been invoked with some success for the very longest transients (see section~\ref{tde}), and
secondly, the first two events with detailed follow-up (GRB 101225A and GRB 111209A) appeared to lie in locations consistent
with the nuclei of their hosts in deep imaging \citep{levan14}. 

However, classical versions of the TDF models in which main sequence stars are disrupted by supermassive black holes
rapidly run into problems. Importantly, the timescale of ULGRBs of $\sim 10^4$ is significantly shorter than
the orbital period at the tidal radius of a $10^6$ M$_{\odot}$ black hole and main sequence star. Secondly, 
the locations of other events, in particular GRB 130925A are clearly non-nuclear, and so would require 
unusual merging systems to create (although it should be noted that in the case of GRB 130925A this
is a distinct possibility). 

A more appealing prospect is that these events do not represent normal TDFs, but rather variants in which lower
mass, intermediate mass black holes tidally shred white dwarfs \citep{krolik11,levan14,macleod14}. 
At these low masses the tidal radius for a white-dwarf can
lie outside the Schwarschild radius, something impossible for more massive black holes. The compact orbit naturally matches
the timescale observed in ULGRBs, and, unlike SMBH systems, it might be expected that IMBH could arise 
in young stellar clusters, where they may also power Ultra- or Hyper-luminous X-ray sources, or in the cores
of globular clusters, all of which would place them well away from the nuclear regions of more massive galaxies, but
perhaps in the nuclei of low mass systems, such as the hosts of GRBs 101225A and 111209A. 

\subsection{Future prospects}
The combination of extremely long durations, with frequently unusual and unexpected multi wavelength
properties offers a strong guide that the ULGRB population does represent a genuinely new class
of burst. 
The crucial question underlying studies of the ULGRB population is then the unveiling of the progenitors. This
diagnostic would both directly answer the question of whether ULGRBs are genuinely a new population, or 
the long tail of the ULGRB distribution, and identify how they might be used as cosmological probes. If they were 
somehow related to SN Ic as normal LGRBs are then it would clearly be important to understand what features 
in the core collapse give rise to such long-lived and luminous central engines. Models that rely to date on fallback to
create plateaus
or on fragmentation of the supernovae \citep{king05}, or in the disc \citep{perna06} would seem to fail to provide the necessary late
time accretion rates, and so our understanding of core-collapse in GRB environments would seem in need of revision. 

Alternatively, should they arise from the collapse of giant stars then they provide a demonstration that 
engines can be active in a far greater range of supernovae than currently appreciated. The ability of these
engines to power GRB-like events in massive stars that retain their hydrogen envelopes may make ULGRBs 
relatively common through the Universe, while their observed rarity is largely due to selection effects \citep{levan14}. 
Indeed, the ability for engines to be active in relatively extended stars may bode well for their use as probes
of population III stars \citep{piro14}. 

Finally, should they in fact arise from tidal disruptions involving intermediate mass black holes then they would 
provide one of the few robust routes to the identification of such systems, and in turn enable uses as probes of
the locations of IMBHs in dwarf galaxies and globular clusters, as a route to hone the $M-\sigma$ relation, and
as an input to models of galaxy formation and evolution. 

It is interesting to note that all of the above systems seem plausible for the origin of ULGRBs, both in the sense
that the rates of the relevant progenitor systems mean they should occur, and because more detailed simulations
manage to create transients with the relevant durations in each case \citep{nakauchi13,macleod14}. It may then be the case that more than
one of the above scenarios provide an explanation for ULGRBs (or at least long duration high energy transients more 
generally), or that there are alternative routes of identifying the other channels. 

The actual task of making the distinction between these models should be relatively straightforward with current
technology. SNe in GRBs can be observed from the ground to $z \sim 1$, and potentially beyond with {\em HST}. This hypothesis is already stretched by the existing observations, but can be rigorously tested with a modest set of new observations. The SNe associated with the collapse of giants should be hydrogen rich, but could be an order of 
magnitude fainter than those seen in GRBs. This makes their detection more difficult, but the presence of strong, variable H$\alpha$ emission should make them traceable out to similar redshifts with current IR spectrographs, or again with {\em HST} observations. Finally, in white dwarf disruptions there may be no SNe, or an unusual type I event (different from the type Ic associated with LGRBs \citep{macleod14}). It is interesting to note that the origins of both LGRBs, and recently of SGRBs have been pinned down by studies of late, red bumps in their optical and IR light curves \citep[e.g.][]{hjorth12,cano,tanvir13b,berger13b}. These studies are likely to be equally diagnostic for ULGRBs, and should be possible in the next few years.

\begin{figure}[h!]
\begin{centering}
\includegraphics[width=\columnwidth]{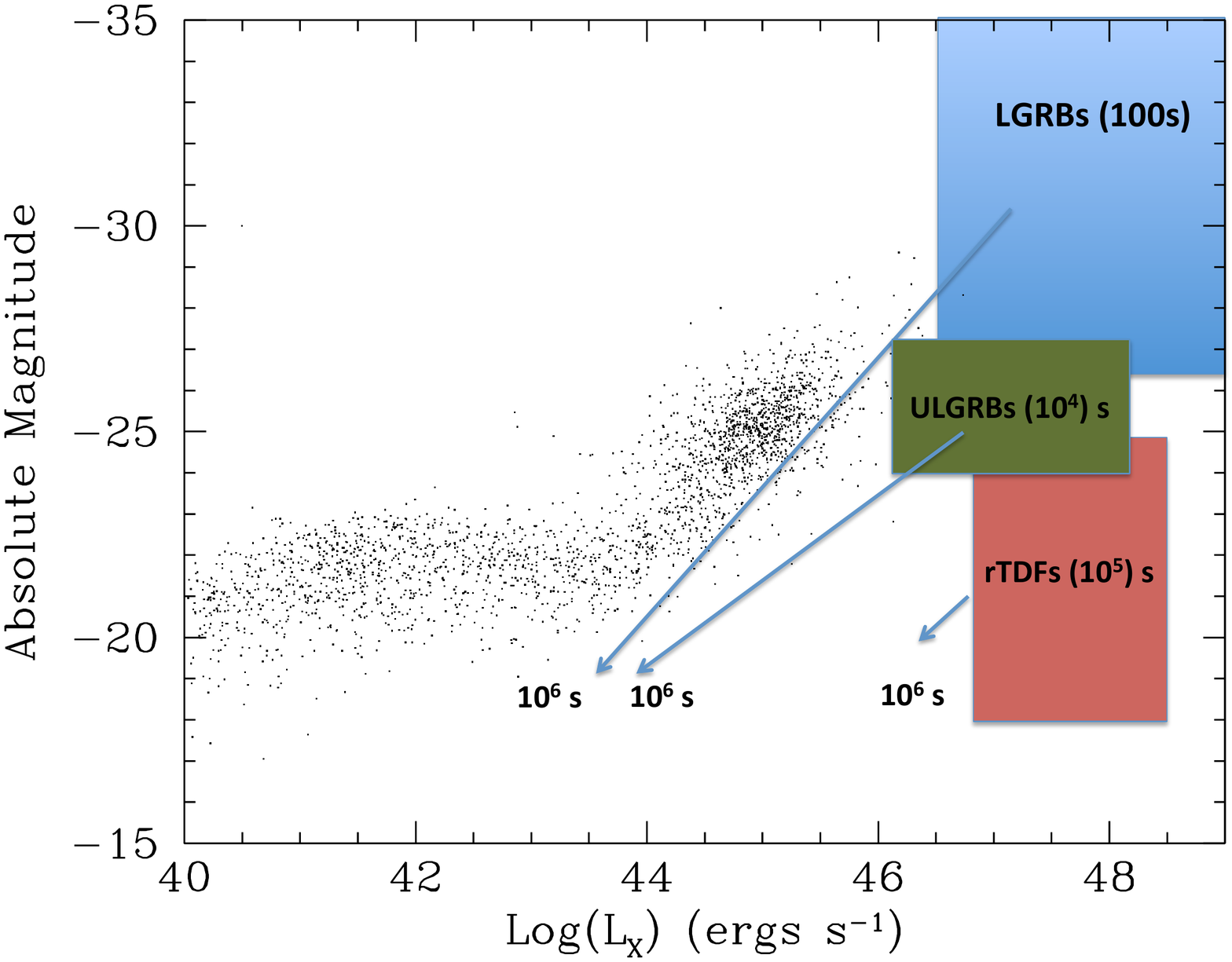}
\vskip -0.05truein
\caption{Cartoon phase space for various classes of high energy transient of differing durations, modified from 
\cite{levan11}. The axes shown an X-ray and optical luminosity (expressed as an absolute magnitude), while the background points arise from
a cross correlation of the 2-XMM catalog \citep{2xmm} with SDSS \citep{sdss}, and a sample of local galaxies observed with {\em XMM-Newton} \citep{ho}. Each population (LGRB), (ULGRB), (rTDF) is shown at a typical peak time (100s for GRBs, $10^4$ s for ULGRBs and $10^5$ s for rTDFs). The arrows indicate the evolution of the events to $10^6$ s. There is significant diversity in the LGRB population, and this may well be represented as well in the overall populations of ULGRBs and rTDFs. Therefore
both the locations of the populations, and their evolution should be viewed as approximate. Nonetheless, it is clear the events do occupy quite different regions of phase space, and diagnostics such as this may be powerful discriminants between differing origins for transients, in particular the TDFs. }
\label{phase_space}
\end{centering}
 \end{figure}
 
\section{The longest events, beyond $10^5$ s}
\label{tde}
As well as the population of ULGRBs, that have durations of $T_{90} \sim 10^4$ s, there is an additional
population of newly recognised $\gamma-$ray transients that are detectable for days after the initial trigger (see Figures~\ref{pspace} and~\ref{phase_space}).
These have been detected initially as GRBs, but also via the BAT transient monitor programme \citep{krimm13}
that reconstructed the image plane for BAT and searches for transients on even longer timescales. 
While some very long $\gamma-$ray outbursts arise from either Galactic binaries, or well known
AGN, this new population do not. Instead they occur at
cosmological distances, reach peak luminosities in excess of $10^{48}$ ergs s$^{-1}$ (the
Eddington limit of a $10^{10}$ M$_{\odot}$ black hole), but are found in galaxies showing
little or no evidence for AGN activity. 

These events are generally ascribed as relativistic variants on tidal disruption flares, in which some small 
fraction of the ejecta is emitted at relativistic speed. This model is compelling, but alternatives have been suggested. Perhaps
unsurprisingly these suggestions are very similar to those discussed for the origin of ULGRBs. In particular this
includes both WD-IMBH disruptions, and long lived accretion in very massive, extended stars.

\subsection{The sample: \\ Swift J1644+5734, Swift J2058+0516, Swift J1112-8238}
The best studied candidates TDF is {\em Swift J1644+5734} \citep{levan11,bloom11,burrows11,zauderer11}, which
was detected via a GRB trigger and had substantial additional follow-up. It was initially thought to be likely
due to a soft, fast X-ray transient with the Galaxy, and indeed similarities were noted with the Galactic
system IGR J16479-4514 \citep{kennea11}. However, redshift measurements from Gemini and the GTC  \citep{levan11} 
demonstrated an extragalactic origin at $z=0.35$, and highlighted the extreme 
luminosity of the event. In total {\em Swift} J1644+5734 triggered the {\em Swift}-BAT on four occasions over 48 hours, and
went on to be detectable by BAT for over a week \citep{levan11,burrows11}. 

The second example located is that of {\em Swift} J2058+0516 \citep{cenko12}, which was detected by the
{\em Swift} transient monitor only 2 months after the detection of {\em Swift} J1644+5734. Follow-up observations
again revealed a cosmological source, this time at $z=1.18$, based purely on absorption lines in Keck spectroscopy
(no emission lines were seen in this, or in X-shooter observations covering the common host lines,\cite{pasham15}). 

The final example to date, {\em Swift} J1112-8238 was not located in real time, but in an archival search
within the BAT transient monitor archive \citep{krimm13,brown15}. This means that real-time follow-up was limited, 
but later observations reveal an X-ray and optical transient coincident with the nucleus of a faint (and hence
low signal to noise) galaxy, likely at $z=0.89$ base on a single emission line identified as the 
O[{\sc ii}] 3727\AA~ doublet \citep{brown15}.

\subsection{X-ray counterparts}
The X-ray counterparts of these transients are all extremely luminous and long-lived. They have peak luminosities around $10^{48}$ ergs s$^{-1}$, and remain brighter than $10^{46}$ ergs s$^{-1}$ for 50-100 days. Indeed, all three have remarkably similar luminosities at around the 50 day mark. It is difficult to define the power-law slope precisely in these cases, since the zero time of the burst is unclear. For example, in the best studied case of {\em Swift} J1644+5734, there were multiple triggers of BAT over several days, but detailed searches revealed the source was actually detected at least
4-days before the first of these \citep{burrows11}. Additionally, the light curve shows pronounced dips, that have been
claimed so show some evidence for periodicity \citep{saxton12}. This makes fitting a non-trivial task, although a simple fit
to the late time light curve of {\em Swift} J1644+5734  \footnote{taken from, http://www.swift.ac.uk/xrt$\_$curves/450158} from $10^6$ to $4 \times 10^7$ syields an index of $t^{-1.65}$.
The case of {\em Swift} J2058+0516 appears to be rather steeper than this, with a decay of around $t^{-2.2}$ \citep{cenko12}. Finally, 
for {\em Swift} J1112-8238, the sparse data provides a limited lever arm but is best fit with a decay around $t^{-1.1}$ \citep{brown15}. These decay rates can in principle be compared to the expectations of tidal disruption models, although recent detailed calculations suggest that the simple 
approaches previously taken may be very different from what will actually be observed, although the
$t^{5/3}$ and $t^{-2.2}$ of {\em Swift} J1644+5734 and {\em Swift} J2058+0516 are remarkably close to
certain scenarios \citep{guillochon13}. 

\begin{figure}[h!]
\begin{centering}
\includegraphics[width=\columnwidth]{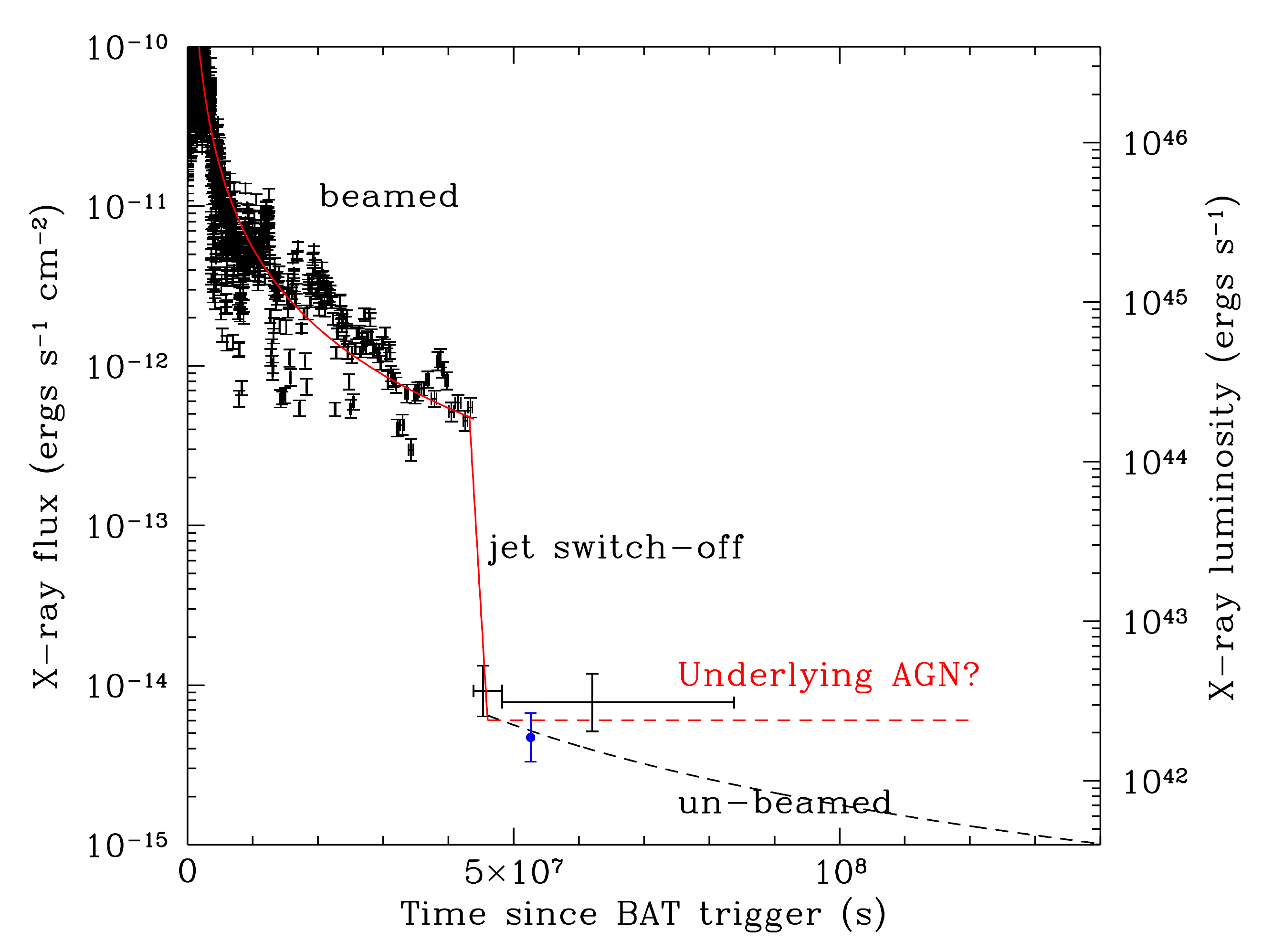}
\vskip -0.05truein
\caption{The X-ray light curve of {\em Swift} J1644+57 on a linear scale, highlighting the rapidity of the switch off. The red line prior
to the switch off is a $t^{-5/3}$ decay, while the extension of this (two orders of magnitude fainter) is shown as the dashed line beneath. At
present it is not clear if the level seen after the drop-off is some continuation of a decay, or a quiescent level in the host galaxy (e.g. from
an underlying AGN). }
\label{tdf_lc}
\end{centering}
 \end{figure}

In addition to this broad decay there are two additional important features in the X-ray light curves. The first is rapid
variability occurring at early and late times, often showing a doubling of the flux on timescales of only a few hundred seconds, even hundreds of days after the burst \citep{levan11,pasham15}. This behaviour is similar to that seen in blazers, and so is likely a hallmark of a relativistic jet. The second striking feature observed in {\em Swift} J1644+5734, {\em Swift} J2058+0516 and possibly {\em Swift} J1112-8238 is a rapid shut off of the X-rays, a year or more after the outburst. in the case of {\em Swift} J1644+5734, the counterpart appeared to switch off within a few days, interpreted as a power-law slope it would have been steeper than $t^{-70}$ \citep{levan15}, see Figure~\ref{tdf_lc}. 

This combination of X-ray properties (luminosity, variability, shut-off) are remarkable, and are not mirrored in any other class of source. 

\subsection{Optical and IR observations}
Optical observations are rather more limited than those in the X-ray for these sources, especially in the case of {\em Swift} J1644+5734, which had nearly daily monitoring at X-ray wavelengths in the months (and even years) since its discovery. However, optical counterparts
have been discovered to all three of the candidate events. These vary in their apparent properties. In the case of {\em Swift} J1644+5734
the source was highly reddened with $A_V \sim 6$, not surprising if it lies in, or close to the host galaxy nucleus \citep{levan11,bloom11,burrows11}. In the case of {\em Swift J1112-8238} only a handful of detections were obtained at the time, since the source was not identified as of interest in real time. Only
for {\em Swift} J2058+0516 do we have a good, and well sampled optical/IR light curve from early to late times after the outburst. This shows clear evidence for a hot, thermal component that cools at near constant radius in the months and years since the outburst. This would appear to be distinct from the non-thermal X-ray emission, perhaps indicative of a different location for its production. However, it is interesting to note that it does apparently switch-off at the same time as the X-ray emission, perhaps indicating a similar origin, despite the rather disparate spectra \citep{pasham15}. 

\begin{figure}[h!]
\begin{centering}
\includegraphics[width=\columnwidth]{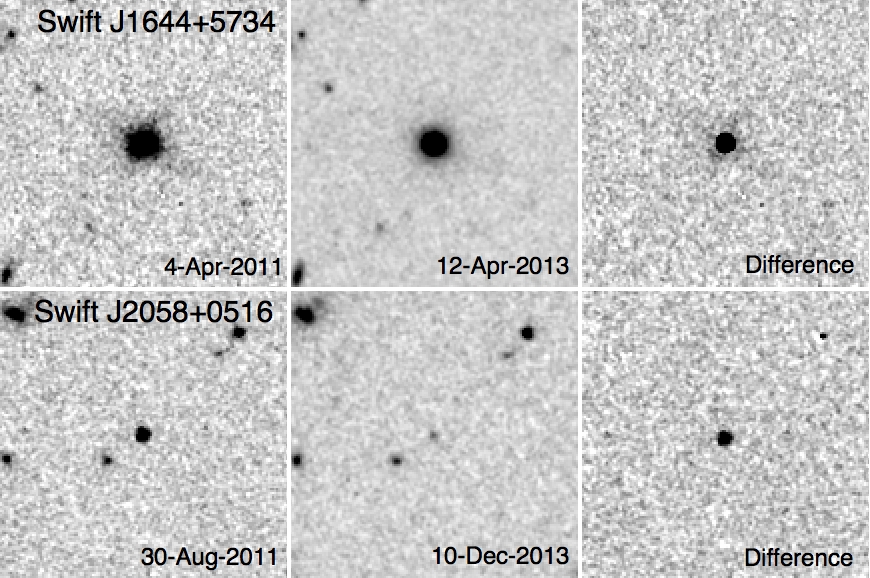}
\vskip -0.05truein
\caption{{\em Hubble Space Telescope} imaging of the host galaxies of {\em Swift} J1644+57 \citep{levan11} and {\em Swift} J2058+0516 \citep{pasham15} in the F160W IR band. The top panel is an early image containing the afterglow and host galaxy light, the second an image showing just the host galaxy, while the third image shows the subtraction of the two epochs. In each case it is clear that the sources lie at locations coincident with the nuclei of their hosts. }
\label{astrometry}
\end{centering}
 \end{figure}

Optical observations provide the strongest possible constraints on the positions of the sources within their host galaxies. In the cases
of {\em Swift} J1644+5734 and {\em Swift} J2058+0516 this is possible thanks to early and late time observations with {\em HST} \citep{levan11,pasham15}, and is shown graphically in Figure~\ref{astrometry}. In these cases the sources are shown to lie
within 150 and 400 pc of the nuclei of their host galaxies in projection, and in each case are consistent with a nuclear origin. In the case of
{\em Swift} J1112-8238 only ground based images are available at the time of writing. These show the source to be associated with
the centroid of its host galaxy, but the low signal to noise precludes strong conclusions from this.

\subsection{Radio counterparts}
One of the most striking features of {\em Swift} J1644+5734 is the presence of a bright and extremely long-lived radio counterpart \citep{zauderer11,berger12,zauderer13}. This rises on a timescale of 200 days after the burst, continuing to brighten well beyond the X-ray peak. Indeed, it is still visible beyond the point at which the X-rays appear to shut-off \citep{zauderer13}. The radio counterpart provides direct evidence for relativistic motion, although with only a modest Lorentz factor $\Gamma \sim 2$ \cite{zauderer11}. Should the jet associated with these outbursts be similar in behaviour to GRB jets this would imply that it
is relatively wide. Limited radio follow-up for {\em Swift} J2058+0516 would seem to match these expectations, although much more sparsely sampled. 

A striking feature of these radio luminosities is that they should be visible to next generation radio surveys to $z>6$. Combined with the longevity of the luminous radio emission (several years at $z \sim 6$) it may well be that these sources will ultimately be best located in the radio regime.

\begin{table*}[htdp]
\begin{center}
\begin{tabular}{lllllll}
\hline
GRB & Trigger & z & peak $L_X$ (erg) & $M_{BH}$ (M$_{\odot}$) &  References  \\ 
\hline 
{\em Swift} J1644+5734 & Burst  & 0.35 & $3 \times 10^{48}$  & $10^5 < M_{BH} <10^7$ &    \cite{levan11,burrows11} \\
{\em Swift} J2058+0516  & Monitor  & 1.19 & $7 \times 10^{47}$  & $10^4 < M_{BH} <3 \times 10^7$ & \cite{cenko12,pasham15} \\ 
{\em Swift} J1112-8238   & Monitor & 0.89   & $7 \times 10^{46}$* & $\sim 2 \times 10^7$ &  \cite{brown15} \\
\hline
\end{tabular}
\end{center}
\caption{The {\em Swift} sample of transients with duration $>10^5$s, the candidate relativistic tidal disruption flares. * indicates that
X-ray observations started very late. The extrapolation back to early times using the BAT transient monitor would have resulting in a peak
luminosity approximately a factor of 10 brighter.  The black hole mass estimates are based on a bulge luminosity to black hole mass relation a at the upper end, and on scaling arguments relating to the variability or rapid shut-off at the low mass end. }
\label{sample_tdf}
\end{table*}%

\subsection{Relativistic tidal disruption flares}
The concept of a tidal disruption flare has long been discussed, and is an inevitable consequence of the
presence of supermassive black holes in galactic nuclei. The disruption occurs when 
a star approaches a black hole sufficiently closely that the gravitational pull on its outer layers 
from the black hole exceeds that from the star itself. At this point the star can disrupt either completely,
or partly (e.g. by stripping of the outer layers of the star). In a complete disruption 
half of the star is placed onto eccentric bound orbits and will return to form a disc and ultimately
accrete onto the black hole, the other half is unbound. The accreting material is shock heated to 
high temperatures peaking in the extreme-UV or soft X-ray regimes \citep[e.g.][]{rees88}. 
Numerous candidates
for such events have been uncovered \citep[e.g.][]{gezari03,komossa04,halpern04,esquej08,saxton12}, although it has been difficult to rule out alternative
explanations, such as unusual nuclear supernovae in many cases. 

However, more recently it was suggested that in addition to this thermal emission 
a small fraction of the material may be expelled at relativistic velocities \citep{metzger,vanvelzen}. Such material would
create a new route to identifying TDFs, and was originally considered in the context of
next generation radio surveys, which could uncover the events at late times when the blast wave
was approximately spherical. 

These extremely long transients provide a natural match to the expectations of this model. There is direct evidence from
the radio observations for relativistic motion, while the super-Eddington accretion rates observed at early times might
suggest an even narrower beam (the early luminosity exceeds the Eddington limit for the black holes within these
host galaxies by a factor of $>10^4$). All of the events are consistent with the nuclei of relatively low mass galaxies, and with black holes with $M_{BH}  < 10^8$ M$_{\odot}$ (see below), consistent
with the disruption of main sequence stars occurring outside the event horizon. Given the theoretical prediction of these events
prior to their detection, their long-lived, high luminous emission, and their locations within host galaxies this model provides
an extremely good fit to the observations, and is adopted as the consensus view for the origin of these outbursts. However, 
it is not a unique model, and other alternative has been put forward as well. 

\subsubsection{Black Hole Masses}
If these events are interpreted as TDFs then various routes have been proposed to measure the black hole masses within them. 
The most straightforward route (though potentially with the largest scatter) is to utilise the well known relationship between
bulge-mass and black hole-mass in galaxies \citep[e.g.][]{haring04}. The total stellar mass of the galaxies can be estimated from rest-frame IR observations, 
and then used to infer a black hole mass. In each case, the masses of the black holes obtained via this route are $M_{BH} \leq 2 \times 10^7$ M$_{\odot}$. 

Alternatively, one can use timing arguments to infer the mass of the black hole. The rapid 
variability of {\em Swift} J1644+5734 implies that any temporally connected source by at
must be smaller than the Schwarschild radius of a $8 \times 10^6$ M$_{\odot}$ BH \citep{bloom11}, providing
a relatively robust upper limit to the black hole mass. 
Alternatively, a rather more complex approach can utilise known relations between radio and X-ray
luminosity, that apparently links stellar and supermassive black holes. This route infers are rather small intermediate mass black hole $\log{M_{BH}} =  5.5 \pm 1.1$ M$_{\odot}$ \cite{miller11}. 

Finally, it has been suggested that the rapid switch off seen in both {\em Swift} J1644+5734 \citep{levan12,zauderer13} and {\em Swift} J2058+0516 \citep{pasham15} might be due to 
the source reaching its Eddington limit. This has some appeal, since at this point it would no longer
be necessary for the source to be beamed in order to drive its luminosity against inflating material. 
In this case, the luminosity at the break would be indicative of the the Eddington limit for that
mass of black hole, providing a route to measuring the mass. A conservative approach would be
to assume that the Eddington limit lies somewhere between the pre- and post-drop luminosity, although
the drops are so steep that this only provides a black hole mass limit in the range $10^4 < M_{BH} < 10^6$ M$_{\odot}$ \citep{pasham15}.

All of these mass estimates contain considerably uncertainty, and rely either on the extrapolation
of known relations significantly beyond the regions in which they have been tested (e.g. the
$M_{\mathrm bulge} - M_{BH}$ relation, or the variability fundamental plane), or on 
hitherto untested relations (i.e. that the jet switch-off occurs at the Eddington limit). However, it is
encouraging that all of these routes yield masses of $M_{BH} < 10^8$ M$_{\odot}$ for all three
events, meaning that they are consistent with the expectations of main sequence stars being
shredded by black holes, comparable in mass to those that we know exist within the local Universe. 

\subsection{White dwarf disruptions}
The early light curve of {\em Swift} J1644+5734 is punctuated by a series of extremely rapid flares, during which the luminosity jumps by orders
of magnitude on a timescale of only a few hundred seconds. These flares, repeat on timescales of around 30,000 s (although it should be stressed that they are not strictly periodic) and have a marked similarity flare to flare. These timescales are perhaps surprising for a main sequence disruption, since they are much shorter than the orbital period of the disrupting star, and therefore consecutive passages of the star (with further disruption) cannot be explanation for the origin of the flares. As an alternative, it has been proposed that
{\em Swift} J1644+5734 is due to a white-dwarf disrupted by a IMBH \citep{krolik11}, an
interestingly similar model to those subsequently proposed by some authors to 
explain the ULGRBs \citep{macleod14}. The appeal of this model lies in the natural interpretation of the flare structure. However, interpreting this structure is challenging as it likely based on the beamed X-ray emission, and so its direct correlation to the $\dot{M}$ onto the black hole is less clear. 
Furthermore, while most models would predict that there is relatively smooth return of material from the disrupted star back to the black hole, this may well not be the case at the point of return of the most bound debris (i.e. the material that first accretes), which could fragment or accrete in sporadic bursts, even from a main sequence star. 

\subsection{Massive star collapse} 
{\em Swift} J1644+5734 was first discovered as a GRB, and its long-lived emission was subsequently interpreted as being related
to a massive star collapse \citep{quataert12}. More detailed models have subsequently been derived by Woosley \& Heger \citep{woosley12}, who demonstrate 
various means by which long-lived transients can arise from collapsar-like events, including single massive stars, tidally locked binaries,
and pair instability collapse. 
Interestingly, these models predict a rapid switch-off in the in accretion rate, essentially when all of the star has been accreted. This may
mirror the switch-off's seen in the events to date, and it is striking that this prediction was made prior to their discovery (see Figure~1 of \cite{quataert12}).  Indeed, further
support that core collapse events can reach such high luminosities comes from X-ray observations of the super luminous supernovae SCP 06F6, that reached $L_X \sim 10^{45}$ erg s$ ^{-1}$, 150 days after its initial outburst \citep{levan13}

However, the locations of the bursts are problematic for massive star collapse models. Even GRBs, that are highly concentrated
on their host like only lie coincident with the nucleus of their hosts $10-15\%$ of the time \citep{fruchter,svensson}. Given this the probability of 
3 events having such a location is $<0.4$\%, apparently ruling out the possibility of a GRB-like origin at $>3\sigma$. However, it should be 
noted that this conclusion is currently only robust in two cases, and deep {\em HST} observations of {\em Swift} J1112-8238 are hence
crucial to this conclusion.

\subsection{Role as probes}
While the origin of the ULGRBs remains an open question, it seems likely (if not yet conclusive) that
the longest events are due to the tidal disruption of stars by massive black holes. These systems
are apparently much rarer (in terms of volumetric rate) than ``classical", non-relativistic tidal flares, which
are expected at rates of $\sim 10^{-5}$ yr$^{-1}$ galaxy$^{-1}$ \citep[e.g.][]{rees88}. 
This is likely because of either
the narrowness of the beamed emission (so that only a small fraction are seen) or because of
the rarity of jets (because few systems launch relativistic jets in the first instance). Crucial questions
therefore arise in the role of these events as probes of jet formation, launching an ubiquity around
massive black holes. The tidal flares are unusual because they provide the opportunity to witness accretion around a supermassive black hole from its start to finish over the timescale of only a few years, therefore providing new constraints about how accretion behaves in a range of regimes. 

Furthermore, since these events may be visible either to X-ray or radio searches to much larger distances than non-relativistic TDFs, it is possible that they may come to dominate the number of
systems observed. If a route can be found to translate their observations into black holes masses this provides the ability to track changes in the black hole mass function through ongoing accretion and heirarchical mergers over most of cosmic history. Indeed, it has been suggested that
merging black holes may raise the tidal disruption rate by 4 orders of magnitude (to one per 10 years or so), in which case their discovery might provide a route of identifying close merger black hole pairs of interest to low frequency gravitational wave detectors (e.g. eLISA) \citep[e.g][]{stone11}.

Finally, the strong particle acceleration in the jets of relativistic tidal disruption flares are promising
sites for the acceleration of ultra-high energy cosmic rays \citep{bloom11,cenko12}. Since their are insufficient bright AGN within the local horizon through which the rays can propagate, and GRBs appear to be insufficient to 
provide this acceleration \citep{icecube12}, it is possible that the tidal flares could provide the solution, although further studies of the details of shock acceleration, and tidal flare rate are clearly necessary to firm this conclusion up \citep{farrar14}.

\section{Open questions and future prospects}
It is remarkable that highly energetic transient events ($E_{iso} >10^{53}$ erg) with durations from hours to days apparently exist, but have been undetected by a fleet of both pointed and survey $\gamma-$ and X-ray telescopes over the past 5 decades. They have evaded detection by a combination of rarity (hence they have not been found serendipitously in the narrow field of view of X-ray telescopes) and low peak-flux (so they haven't, in general, been identified by rate based $\gamma$-ray detectors). These events are only now uncovered in sufficient numbers to identify new populations of outburst, thanks to the ability of {\em Swift} to both locate them, but also rapidly highlight their interest to the community in a timely way. However, these new populations are a recent development, and our knowledge and understanding is much less mature than for the LGRBs and SGRBs, whose origin and physics have been hotly discussed for several decades. Future observations will continue to build on our knowledge of these populations, and address several central questions, including.

\begin{itemize}
\item
Are ULGRBs are separate class, spawned from giant stars, or WD-IMBH disruptions, or do they form the tail end of the LGRB distribution? 

\item
Can a larger sample of candidate relativistic TDFs confirm that they are all related to tidal flares, and not other possible progenitors? 

\item
How large are the true volumetric rates of the systems? 

\item
What are their roles as probes of both cosmology (pop III-like ULGRBs, ubiquity of black holes in galaxies) and of extreme physics
(can their engines explain the origin of ultra-high energy cosmic rays? What do their engines tell us about jet-formation in AGN?). 
\end{itemize}

The study of these extreme, but still mysterious transients remains an open question in astrophysics, it will be at the forefront of work as
{\em Swift} enters its second decade of operations.

\section*{Acknowledgements}
I thank the referee for constructive comments that improved the paper.
I acknowledge the use of the Swift data science centre at the University of Leicester, and thank N. Tanvir, A. Fruchter, G. Brown and S.B. Cenko for useful discussions. This work is supported by STFC under grant number ST/L000733/1 and the Leverhulme Trust via a Philip Leverhulme Prize (2011). 




\end{document}